## Main Manuscript for

# Hidden high-risky states identification from routine urban traffic

Shiyan Liu[1], Mingyang Bai[1], Shengmin Guo[2], Jianxi Gao[3], Huijun Sun[4], Ziyou Gao[4*], Daqing Li[1*]

[1]School of Reliability and Systems Engineering, Beihang University, Beijing 100191, China
[2]State Key Laboratory of Software Development Environment, Beihang University, Beijing 100191, China
[3]Department of Computer Science, Rensselaer Polytechnic Institute, Troy, NY 12180, USA
[4]School of Systems Science, Beijing Jiaotong University, Beijing, 100044, China

* Correspondence should be addressed to: Ziyou Gao; Daqing Li
**Email:** zygao@bjtu.edu.cn (Z. Gao); daqingl@buaa.edu.cn (D. Li)

**Author Contributions:** S.L., M.B. and D.L. designed the research. S.L., M.B., S.G., J.G., H.S., Z.G., and D.L. developed the method. S.L., M.B. performed and checked the experiments. All the authors analyzed the results. S.L. and D.L. wrote the paper with comments from all the authors.

**Competing Interest Statement:** The authors declare no conflict of interest.

**This PDF file includes:**

Main Text
Figures 1 to 4

**Abstract**

One of the core risk management tasks is to identify hidden high-risky states that may lead to system breakdown, which can provide valuable early warning knowledge. However, due to high dimensionality and nonlinear interaction embedded in large-scale complex systems like urban traffic, it remains challenging to identify hidden high-risky states from huge system state space where over 99% of possible system states are not yet visited in empirical data. Based on maximum entropy model, we infer the underlying interaction network from complicated dynamical processes of urban traffic, and construct system energy landscape. In this way, we can locate hidden high-risky states that have never been observed from real data. These states can serve as risk signals with high probability of entering hazardous minima in energy landscape, which lead to huge recovery cost. Our finding might provide insights for complex system risk management.

**Significance Statement**

Hidden high-risky states are lethal system states with catastrophic consequences for large-scale complex systems. Existing methods for identifying hidden high-risky states face challenges in routine urban traffic, where over 99% of possible system states are not yet visited in empirical data. Based on large-scale traffic-speed data, we build a least-structured network entropy model that captures intricate traffic interactions. This model can generate the entire state space and help the identification of hidden high-risky states that may lead to extreme events. Our results suggest a new way to provide reliable early warning signals for system breakdown based on critical risk scenarios.

## Main Text

### Introduction

Hidden high-risky states are system states that have not occurred yet but are likely to happen in the future with catastrophic consequences. Once these states occur, they may lead to system breakdown[1] in natural, technical and social systems, such as earthquakes[2], tsunamis[3], financial crises[4], blackout[5], extreme climate[6], neuronal avalanches[7], etc. In urban traffic, system breakdown is generally understood as conditions in which traffic is completely or large-scale paralyzed, leading to enormous damage. For example, on July 21, 2012, Beijing was hit by violent rainstorm that paralyzed the backbone of road network[8]. On May 23, 2014, São Paulo suffered the worst traffic jam in its history with 214 miles of congestion around the city due to the opening ceremony World Cup and heavy rains[9]. In future traffic management scenarios with more connected vehicles on the road, cyberattacks against smart traffic management systems may lead to widespread traffic paralysis[10]. To prevent these system breakdowns from occurring, identification of hidden high-risky states is needed. However, state space of urban traffic system is so huge that traditional methods can hardly recognize a small number of states that leads to breakdown. One of the essential questions is how to effectively identify these hidden high-risky states that serve as precursors to system breakdown in urban traffic.

To effectively identify the hidden high-risky states that may lead to system breakdown, researchers have dedicated extensive efforts. There are several traditional methods for identifying system breakdown which mainly focus on their statistics and indicators, such as extreme value theory[11], large deviation theory[12], etc. To further address this challenge, a wide range of advanced methods have been developed, which can be broadly classified into data-driven methods or model-driven methods. In the first route, historical spatiotemporal observations are used to determine whether and when system breakdown will occur without knowing system dynamic models. Examples are forecasting extreme floods using an event synchronization measure[13], global seasonal forecasts of marine heatwaves by a large multi-model ensemble[14], system breakdown prediction via active learning[15] and deep learning-based methods[16,17]. The second route is based upon the identified dynamic models of systems to uncover breakdown mechanisms, like bursting phenomena of multiple time scales in deterministic systems[18,19], and noise-induced transitions to extreme state attractor in non-deterministic systems[20-22]. In above approaches, data-driven methods often require extensive sample data, while the highly cyclical and repetitive nature of daily traffic patterns[23] results in an exceptionally low occurrence of system breakdown. Meanwhile, model-driven approaches typically face challenges in complex systems like urban traffic which are controlled by a large family of hidden correlated parameters. To overcome these issues, we attempt to build a least-structured network model that captures intricate traffic interactions, which enables reconstruction of the entire state space and identification of hidden high-risky states.

Specifically, we use the spatiotemporal data of urban road network to build a pairwise maximum entropy model[24-28], through which we infer the underlying interaction network between local regions. Based on this model, we define hidden high-risky states and identify them by constructing system energy landscape and analyzing their cascading processes. Our proposed method can identify hidden high-risky states from routine urban traffic, providing a list of potential management targets often overlooked in traffic management practices. In addition, the model is built from the perspective of interactions between local regions rather than road segments (reducing the system state space from $2^{33,000}$ to $2^{20}$), which avoids system dimensionality explosion. This makes it possible to identify hidden high-risky states more effectively and analyze their dynamic properties on the whole state space. Our results suggest a new way to provide reliable early warning signals for system breakdown and reconstructed critical risk scenarios for traffic management practices.

## Results

### Maximum entropy model for urban traffic

Here, we consider a specific local road network area as a hexagonal region, as shown in Fig. 1a. Overall road network states are represented by binary states of 20 regions (i.e., free region or jam region, see Methods) with their interactions, giving a total of $2^{20}$ possible system states. As a large-scale and high-dimensional complex system, there are a huge number of states within the state space of urban traffic. Despite advances in traffic big data technology that enable real-time monitoring of urban traffic states, highly repetitive nature[23] of daily traffic means that only a small portion of the state space can be observed. We find in real data that the observed states (Fig. 1b and Fig. 1c) of 1-month working days within the system state space are less than $10^3$ (see Methods), 0.1% of the whole state space. This implies that over 99% of the possible states are not yet visited in empirical data. Therefore, by real-time data alone, it is nearly impossible to detect all states of the entire road network[29], especially making identification of hidden high-risky states challenging. To effectively identify hidden high-risky states of urban traffic, a comprehensive understanding for the whole state space is required, the core of which is to infer the dynamical model capable of capturing complex system interactions.

To address this issue, we develop a pairwise maximum entropy model[24] for urban traffic based on real data. Pairwise maximum entropy model is a probability distribution model (see Methods) that can capture interaction network between different local road network regions (Fig. 1d). Model parameters are determined by maximizing (Gibbs) entropy (i.e., $H$) of the joint probability distribution, subject to normalization condition and the constraints of first and second moments (see Methods for more details):

$$\text{Max} \quad H = -\sum_{k \in n} p(s_k) \log p(s_k), \tag{1}$$

where $n$ is number of network states in entire state space, $p(s_k)$ is the probability of network state $s_k$. This probability distribution can be obtained by solving Lagrange function[30] (Supplementary Note 2) as:

$$p(s_k) = (1/Z)\, \mathrm{e}^{\sum_{i \in m} h_i s_k^{(i)} + \sum_{i,j \in m; i<j} J_{ij} s_k^{(i)} s_k^{(j)}}, \tag{2}$$

where $m$ is number of regions, and $s_k^{(i)}$ is state of region $i$ with value +1 (i.e., jam region) or -1 (i.e., free region) when system is in $s_k$, and $Z$ is the partition function, equaling to $\sum_{k \in n} \mathrm{e}^{\sum_{i \in m} h_i s_k^{(i)} + \sum_{i,j \in m; i<j} J_{ij} s_k^{(i)} s_k^{(j)}}$.

The pairwise maximum entropy model is least structured model incorporating second-order interaction terms, the form of which can be mapped onto the Ising model[25,31]. The model parameter $h_i$ represents jam tendency of local road network region $i$, while $J_{ij}$ characterizes the interaction between region $i$ and $j$. Specifically, positive $J_{ij}$ value means a positive interaction[25,31]. This implies that congestion in one region is likely to trigger congestion in the other. Conversely, negative $J_{ij}$ value represent a negative interaction[25,31], suggesting that congestion in one region can alleviate congestion in the other. These model parameters (i.e., $h_i$ and $J_{ij}$) can be estimated from real data (see Methods and Supplementary Fig. 1). We then use $J_{ij}$ to infer underlying interaction network between local regions (Fig. 1d), which exhibits high heterogeneity. For example, region 20 tends to have more positive interactions with other regions, while region 15 tends to have more negative interactions. The heterogeneity embedded in underlying interaction network of urban traffic can lead to highly complicated dynamic behavior[32].

The pairwise maximum entropy model constructed for urban traffic has good performance. At the microscopic level, two moments constraints derived from the maximum entropy model match well with actual data (Fig. 1e), which is unaffected by change of time periods (Supplementary Fig. 2a) and variable thresholds (Supplementary Fig. 2c). At the macroscopic level, the distribution of traffic functional performance obtained from the model retains its distribution properties in the data. Here, traffic functional performance ($G$) is defined as ratio of largest functional cluster[33] formed by connected free regions to the total number of regions. The model preserves the bimodal distribution of $G$ in actual data during rush hours (7:30am~8:30am) (Fig. 1e) as well as the single-peaked distribution of $G$ during non-rush hours (6:00am~7:00am) (Supplementary Fig. 2b). Additionally, we observe a relatively consistent distribution of macro-level road network performance before and after partitioning the entire road network into regions (Supplementary Fig. 3), indicating the feasibility of the regional modeling approach applied to the urban traffic. Unlike traditional correlation methods including Pearson correlation coefficient, our pairwise maximum entropy model considers global system behavior as a result of inferred interaction network embedded in urban traffic, rather than just linear relationships between pairs of regions. This ensures the consistency of global system properties (e.g., the distribution of traffic functional performance) between model and actual data, thereby better capturing the intrinsic properties of urban traffic.

**Hidden high-risky states defined in system state transition network**

To clarify the definition of hidden high-risky states, we first construct entire system state space of urban traffic, which can be represented by state transition network. As shown in Fig. 2a, a node in the state transition network represents a network state $s_k$ ($k$=1,2,3…, $2^{20}$), with edges representing state transition (i.e., one-region-state-flip). According to traffic functional performance of network states ($G$), network states can be classified into normal states with $G \geq 0.5$ (e.g., $s_1$, $s_4$ and $s_5$ in Fig. 2a) or hazardous states with $G<0.5$ (e.g., $s_2$, $s_3$ and $s_6$ in Fig. 2a). Among them, some states (e.g., $s_3$, $s_4$ and $s_5$ in Fig. 2a) have never been observed in empirical data (i.e., hidden states), making them potential risky states that traffic managers may overlook. When urban traffic is in normal states (e.g., $s_5$ in Fig. 2a) with also many normal neighboring states, there is a high probability that it will remain normal, representing stable healthy system operation. In contrast, normal states (e.g., $s_4$ in Fig. 2a) surrounded by many hazardous states are highly likely to fall into hazardous states, signaling that the system is at risk boundary from hazardous states. Regarding hazardous states, those like $s_6$ in Fig. 2a with normal neighbors may spontaneously recover normal, reflecting system resilience. Hazardous states (e.g., $s_3$ in Fig. 2a) with neighboring hazardous states are likely to remain in hazardous states, suggesting relatively stable hazardous condition that could result in significant recovery cost for the system. Here, we aim to identify hidden high-risky states that currently appear normal but with high probability of transition into relatively stable hazardous states. This dynamic characteristic of hidden high-risky states requires early intervention by traffic managers or smart control center to prevent system from persisting in poor traffic operational connectivity. The key question is how to explore dynamical transition processes of hidden high-risky states.

Based on the maximum entropy model of urban traffic, energy of all system states can be calculated, which is a crucial parameter in describing system dynamical behavior. Similar to the definition of energy in Ising model, here energy value $E(s_k)$ of each network state $s_k$ can be calculated by parameters of the learned pairwise maximum entropy model (i.e., $h_i$ and $J_{ij}$) as:

$$E(s_k) = \sum -h_i s_k^{(i)} - \sum J_{ij} s_k^{(i)} s_k^{(j)} . \quad (3)$$

According to (2), the energy value $E(s_k)$ of network state $s_k$ represents its probability of occurrence $p(s_k)$:

$$p(s_k) \propto \mathrm{e}^{-E(s_k)} , \quad (4)$$

indicating that a network state with lower energy value has higher probability of occurrence. It can be seen from (4) that the occurrence probability distribution of network states has the same form as the Boltzmann distribution, which means that energy of each network state can be obtained by counting frequency. However, observed states that account for less than 0.1% are quite sparse, making it impractical to calculate the energy of all network states directly from the actual data alone. Hence, the learned pairwise maximum entropy model is needed to get the energy of states in the whole state space. More importantly, as the system tends to transition from high energy state to lower energy state, the energy difference between two adjacent network states determines their transition direction.

To explore energy characteristics of system states, we analyze energy distribution across all states within urban traffic state space and find that there are a huge number of hidden states with high energy. This explains the observed rarity of these states in real data. As shown in Fig. 2b, energy of all network states is distributed over a wide range between -20 and 30. While a few observed states concentrate around -15, there are still a fraction of hidden states with low energy that have never been observed in real data. For example, it is shown in Fig. 2b that there are 8,877 network states with energy levels below -14, among which only 570 are observed states, indicating that over 93% of low-energy network states remain unobserved in realistic operation. Compared to hidden states with higher energy, hidden states with lower energy have a greater occurrence probability according to (4). As a result, we can filter hidden high-risky states from low-energy states (i.e., high-p states), with energy lower than a certain threshold ($E^{th}$).

By combining above dynamic characteristic with occurrence probability constraints, we define hidden high-risky states as unobserved normal states with high likelihood of occurring and falling into hazardous states (see Methods). It is shown in Fig. 2c that high-p states (i.e., states with high likelihood of occurring) consist of 17,121 network states, about 1.6% of total states, distributed over different values of $G$. Since states usually move from high to low energy, hidden high-risky states will flow into lower energy states, thus becoming stable in low-energy zone. Therefore, for simplification, we will focus on the transition dynamics of low-energy states (i.e., high-p states) rather than the overall states to further identify hidden high-risky states.

**Vulnerability origin from system energy landscape construction**

To quantify transition dynamics of system states, we construct energy landscape of urban traffic, characterized by system state transition network and energy values of system states (see Methods). Energy landscape approach has been successfully used to understand the dynamic process of complex system such as specific folding pathways of proteins[34] and dynamic brain activities during bi-stable perception[35]. In our constructed energy landscape, states move along energy-decrease-paths until reaching local minima, corresponding to local lower-energy states in Fig. 3a. Such local energy minima are critical in various fields such as differentiated cells[36], native state of a protein[34] and stable brain states[35] due to its stability during dynamical processes. This dynamical stability may link to stable hazardous states, which generates high risk. To further analyze these local energy minima, we construct a disconnectivity graph[37] (Fig. 3b) that can be used to describe energy levels of minima and energy barriers between them (see Methods). We find that there are seven local energy minima, representing metastable states of urban traffic[38]. This suggests that city traffic exhibits a rugged energy landscape similar to certain type of spin glass[39]. This may be attributed to the frustration[40] arising from complex positive and negative interactions (i.e., $J_{ij}$ in pairwise maximum entropy model) between local regions.

Unlike numerous valleys in the energy landscape of two-dimensional spin glass, urban traffic landscape contains only seven valleys, given its high-dimensional landscape[41].

Among these seven local minima, there are two normal minima and five hazardous minima; three observed minima and four hidden minima. Accessibility of each minimum is analyzed based on its basin size, which is the number of states that can reach it. Compared to normal minima, accessibility of hazardous minima is relatively higher. As shown in Fig. 3c, basin sizes of normal minima 3 and 4 are smaller than those of hazardous minima 1 and 2, and basin size of normal minimum 4 is also smaller than that of hazardous minimum 5. The higher accessibility of hazardous minima indicates great vulnerability in the urban traffic. It is worth noting that hazardous minimum 5 in Fig. 3c is unobserved but highly accessible, which represents hidden hazardous state that system likely falls into, revealing potential vulnerability of urban traffic. To further explore system intrinsic vulnerability behavior, we analyze the transitions between two types of minima (i.e., normal or hazardous). Compared to distances between other hazardous minima, the distances between minima 1 and 2, as well as between minima 2 and 5, are shorter. Specifically, they can be reached by changing the states of only three regions (Fig. 3d). With shorter distances, minima 1, 2 and 5 in Fig. 3b tend to form a larger vulnerability valley. Once system enters this larger vulnerable valley, returning to normal requires a high cost, highlighting high risk of urban traffic (during rush hours).

**Hidden high-risky states identification**

Hidden high-risky states are unobserved normal states with high likelihood of occurring and falling into hazardous states. Since hidden normal states (Fig. 2a) and high-p states (Fig. 2c) have been clarified, the dynamic characteristic is the key part of hidden high-risky states identification. Compared to states that only reach normal minima (i.e., class2 in Fig. 4a), states that can reach hazardous minima (i.e., class1 and class3 in Fig. 4a) are more likely to fall into hazardous states. Therefore, the dynamic characteristic of hidden high-risky states is quantified by introducing a risk-level indicator ($R$; see Methods), where a high $R$-value indicates short path-length towards hazardous minima, suggesting great risk.

It is shown in Fig. 4b that there is a broad distribution of $R$-value ranging from 0.01 to 100 for all normal high-p states, indicating significant differences in their risk levels. To verify that this proposed risk-level indicator can measure dynamic characteristic of system falling into hazardous states, we analyze the actual evolution processes of observed high-p states at different values of $R$. Compared to small-$R$ (i.e., $R<1$), normal high-p states with large-$R$ (i.e., $R\geq10$) are more likely to transition into hazardous states in the real operations. As shown in Fig. 4c, as length of time window increases, the probability of system transitioning into hazardous states from observed normal high-p states with large-$R$ increases rapidly, approaching 60% from 35% within 15 minutes and 70% within 30 minutes. Conversely, the likelihood of observed normal high-p states with small-$R$ falling into hazardous states keeps below 45%. This result indicates that even though normal high-p states with large-$R$ seem to have good traffic functional connectivity, they tend to fall into hazardous states during dynamical processes, thus becoming high-risky states. Among these high-risky states, the unobserved states are hidden high-risky states we aim to detect. Traffic managers or smart control center may easily ignore hidden high-risky states because they show good traffic operational connectivity and have not been observed in the data yet. However, without timely intervention in these high-risky states, system will fall into hazardous states with high probability.

To further clarify the consequence of system falling into hazardous states (i.e., $G<0.5$), we use the risk-level indicator (i.e., $R$) to analyze dynamic characteristic of hazardous high-p states. As shown in Fig. 4d, $R$-value of high-p states with $G<0.3$ is quite different, ranging from 0.1 to 100. High-p states with $G<0.3$ and large-$R$ (i.e., $R \geq 10$) persist in hazardous states due to their proximity to hazardous minima (Fig. 4e), demonstrating stable hazardous dynamics. Conversely, high-p states with $G<0.3$ and small-$R$ (i.e., $R<1$) are closer to normal minima, which can escape from such hazardous states spontaneously (Fig. 4e), highlighting system resilience. In high-p states with $G<0.3$, only 30.4% have small $R$-value (i.e., $R<1$), indicating that most high-p states with $G<0.3$ are unable to recover spontaneously. As a result, once system falls into such hazardous states, it is likely to keep hazardous, leading to system breakdowns. Therefore, our identified hidden high-risky states can offer reliable early warning signals for system entering potential stable hazardous states and enhance urban traffic management resilience. We further analyze spatial characteristic of Top 10 hidden high-risky states with the largest $R$-value (Supplementary Fig. 6a), and find that jam regions in these states form several larger jam clusters, aggregation of which may trigger cascading failures[42,43]. In contrast, jam regions of Top 10 hidden normal high-p states with the smallest $R$-value (Supplementary Fig. 6b) are relatively dispersed, making it less likely to form a large congestion cluster. Therefore, avoiding concentration of congested areas may be a primary goal in managing hidden high-risky states.

**Discussion**

This study introduces a novel approach for identifying hidden high-risky states of routine urban traffic. A pairwise maximum entropy model is constructed from real spatiotemporal traffic data to infer underlying interaction network that can capture risk dynamics. This model considers the overall probability distribution of global network states rather than just linear relationships between pairs of regions like traditional correlation methods. Based on this model, we define hidden high-risky states and construct the energy landscape to identify them. Specifically, hidden high-risky states are defined as unobserved normal states with high likelihood of occurring and falling into hazardous states. Hidden high-risky states are identified by their transition processes towards hazardous minima in the constructed energy landscape.

Our identified dynamical processes of high-risky states from normal state to hazardous states is actually phase transition[44], which is widely found in complex systems. It is challenging to discover early warning signals for phase transition. There are efforts including critical slowing down near tipping points[6], changing stability landscapes in stochastic systems[45], etc., and relevant metrics including recovery rate[46] and autocorrelation[47] are proposed. Different from these methods that focus on macroscopic indicators, our identified high-risky states could be good spatio-temporal signals for phase transition in urban traffic. On this basis, resilience regulation of traffic congestion at different transition stages is possible. When traffic system enters states that may fall into hazardous attractors, timely intervention is required to avoid greater losses. Together with smart control agent for urban traffic[48], it is possible to recover system with high resilience.

Our results can also serve for reliability testing scenarios of various smart control agents. With rapid progress of new generation information technologies (e.g., cloud computing, big data and artificial intelligence), smart traffic control agents (e.g., Intelligent Transportation Systems[49], City Traffic Brain and Smart Traffic Signal Systems) have been widely applied, aiming at efficient, intelligent, and sustainable control over the entire road network state. Currently, a key challenge for these smart control agents is to find critical scenarios to accelerate their large-scale reliability testing. Based on our constructed energy landscape, risky states can be sampled as critical scenarios, which may help accelerate reliability assessing and testing of various smart control agents in urban traffic.

## Methods

### Datasets description

Our dataset includes the real time velocity records of each road segment in the Beijing road network, spanning rush hours (7:30am~8:30am) and non-rush hours (6:00am~7:00am) of 17 working days in October 2015. The spatial scope consists of the road network within the Fourth Ring Road of Beijing, covering over 18,000 nodes and more than 33,000 edges. Road speed data is derived from the global positioning system (GPS) recorded in floating cars at 1-minute resolution.

### Definition for local region state and global network state

**Local region state.** In our regional modeling for Beijing road network (see Supplementary Note 1), each local region state can be determined through following steps. (1) Identify congested road segments for each region under given congestion ratio of road segments ($f$) and time $t$. (2) Calculate operational performance for each region at time $t$, $q^i(t)$ ($i$=1,2,3…,20), which is defined as the maximum size of connected cluster formed by congested road segments within it. A local region with a larger $q$-value indicates that it is in a worse operational performance, i.e., vehicles in this region travel at relatively low speeds. (3) Obtain the binary state for each region at time $t$, $s^i(t)$ ($i$=1,2,3…,20). Given region state threshold $q^{th}$, if $q^i(t)$ of region $i$ ($i$=1,2,3…,20) exceeds $q^{th}$, region $i$ at time $t$ is considered jam state, i.e., $s^i(t)$ is set to 1; otherwise, it is regarded as free state, i.e., $s^i(t)$ is set to -1.

**Network state.** Network state is represented as a set of 20-dimensional vectors determined by binary states (i.e., free or jam) of 20 regions. Thus, total number of network states in system state space is $2^{20}$. For a given congestion ratio of road segments ($f$) and region state threshold ($q^{th}$), the network state $s(t)$ at time $t$ can be determined in actual data, i.e., $s(t)=\{s^1(t), s^2(t),…, s^i(t),…, s^{20}(t)\}$ with $s^i(t)\in\{1,-1\}$. Specifically, here we set $f$ to 0.25 and $q^{th}$ to 0.09 to ensure feasibility of regional modeling approach (Supplementary Fig. 3), and record network states at each minute during 17 working days of October 2015. Results show that the number of network states observed in the data during rush-hours (7:30am~8:30am) is only 910, and 362 during non-rush hours (6:00am~7:00am). Compared to entire system space, the number of observed network states in the data is tiny.

### Pairwise maximum entropy model

**Model Introduction.** Maximum entropy model is a modeling approach that uses the maximum entropy principle[24] to find a probability distribution based on given information. If our knowledge of a system is limited, we should choose a probability distribution that maximizes system uncertainty under those constraints. This is least structured or most unbiased model among all probability distributions satisfying certain given constraints. In this context, pairwise maximum entropy model maximizes the entropy of the probability distribution under two moments constraints (i.e., attributes of both individual variables and pairwise correlations between variables match with the actual data). Pairwise maximum entropy model has been widely applied in complex systems to addresses inverse statistical problems[30], such as finding weak interactions between neurons behind strongly collective behavior of neurons[25], deducing the network structure of brain functional areas[27], and analyzing the gene-gene interaction network that underlies cell metabolism[28]. The core of these problems is to infer interaction relationships between local entities from system behavior data while preserving some macroscopic behavioral characteristics of the system.

**Pairwise maximum entropy model of urban traffic.** Pairwise maximum entropy model of urban traffic is a probability distribution model that can capture interaction network between different local road network regions. Model parameters are determined by maximizing (Gibbs) entropy of the joint probability distribution, subject to normalization condition and the constraints of first and second moments:

$$
\begin{aligned}
\text{Max} \quad & H = -\sum_{k\in n} p(s_k)\log p(s_k) \\
\text{s.t.} \quad & \sum_{k\in n} p(s_k) = 1 \\
& \sum_{k\in n} p(s_k) s_k^{(i)} = \left\langle s^{(i)} \right\rangle_{data}, \quad i \in m \\
& \sum_{k\in n} p(s_k) s_k^{(i)} s_k^{(j)} = \left\langle s^{(i)} s^{(j)} \right\rangle_{data}, \quad i, j \in m
\end{aligned}
\quad , \tag{5}
$$

where *n* is number of network states in entire state space, $p(s_k)$ is the probability of network state $s_k$, and $s_k^{(i)}$ is state of region *i* with value +1 (i.e., jam region) or -1 (i.e., free region) when system is in $s_k$. *m* is number of regions, and $\sum_{k\in n} p(s_k) s_k^{(i)}$ is expected state of region *i* and $\sum_{k\in n} p(s_k) s_k^{(i)} s_k^{(j)}$ is expected pairwise joint state of region *i* and region *j*, which can be calculated by estimated probability distribution. $\left\langle s^{(i)} \right\rangle_{data}$ is average state of region *i* in empirical data, which is given by $\left\langle s^{(i)} \right\rangle_{data} = (1/n_D)\sum_{d\in D} s_d^{(i)}$, with larger values representing more congested regions. $\left\langle s^{(i)} s^{(j)} \right\rangle_{data}$ is average pairwise joint state of region *i* and region *j* in empirical data, which is given by $\left\langle s^{(i)} s^{(j)} \right\rangle_{data} = (1/n_D)\sum_{d\in D} s_d^{(i)} s_d^{(j)}$, with larger values representing higher correlation of two regions. Here $n_D$ represents the size of sample dataset *D* and $s_d^{(i)}$ represents state of region *i* in sample state *d*.

**Model Algorithm.** We use Minimum Probability Flow (MPF) algorithm[50] to estimate parameters ($J_{ij}$ and $h_i$) in the pairwise maximum model (i.e., coniii[51]). Specifically, after determining network state at each time point during a given period, model parameters can be estimated based on the frequency of these observed states. We analyze the fitting performance of model through two ways: (1) whether the model satisfies two moments constraints, and (2) whether the distribution of traffic functional performance (i.e., $G$) obtained by the model is consistent with the data.

**Definition of hidden high-risky states.**

We define hidden high-risky states as unobserved normal states with high likelihood of occurring and falling into hazardous states. According to this definition, hidden high-risky states, $S^{\mathrm{HHR}}$, can be represented as:

$$S^{\mathrm{HHR}} = S^{\mathrm{U}} \cap S^{\mathrm{N}} \cap S^{\mathrm{HP}} \cap S^{\rightarrow \mathrm{SH}}, \quad (6)$$

where $S^{\mathrm{U}}$ represents hidden states that have never been observed, and $S^{\mathrm{N}}$ represents normal states (i.e., $G \geq 0.5$), and $S^{\mathrm{HP}}$ represents low-energy states (i.e., high-p states), with energy lower than a certain threshold ($E^{\mathrm{th}}$). Here, we set the energy threshold ($E^{\mathrm{th}}$) to -13.2365, corresponding to low-energy states with a probability of occurrence exceeding $1 \times 10^{-5}$. $S^{\rightarrow \mathrm{SH}}$ represents states that are likely to fall into hazardous states.

**Dynamical processes of network states**

**Energy landscape.** Energy landscape of urban traffic is characterized by system state transition network and energy values of high-p states (Fig. 2c). The dynamics of this energy landscape are defined such that states can only transition to neighbors (i.e., differing at only one region state) with lower energy. Consequently, states move along energy-decrease-paths in the energy landscape until reaching a local minimum. Basin size of each local minimum is determined by number of states that can reach it, thereby representing its accessibility. To emphasize these local energy minima, visualization of energy landscape (Fig. 3a) simplifies the dynamics of system states (i.e., a state can only transition to its lowest-energy neighbor), ensuring that each state can only reach one local minimum along the steepest energy-decrease path. The relative position inside each basin is only for illustration.

**Disconnectivity graph.** Disconnectivity graph is usually used to analyze energy landscape of high-dimension complex systems. In a disconnectivity graph, each endpoint on a branch represents a local energy minimum. The height of these endpoints reflects their energy levels, with lower endpoints indicating lower energy. Energy barrier is defined as energy difference that must be overcome for system to transition from one minimum to another, which equals the vertical distance from the starting minimum up to the connecting saddle point (i.e., point where branches intersect between these two minima).

**Risk-level indicator of system states**. To quantify risk level of system states, we propose an indicator, $R(s_k)$, based on the relative path-length of states transitioning towards two types of high-accessibility minima. Here, high-accessibility minima are minima with large basin sizes, i.e. minima 1, 2, 3, 4 and 5 in Fig. 3b and Fig. 3c. $R(s_k)$ can be denoted as:

$$R(s_k) = l^{\rightarrow Normal}(s_k) / l^{\rightarrow Hazardous}(s_k), \quad (7)$$

where $l^{\rightarrow Normal}(s_k)$ represents the shortest path length from state $s_k$ to normal minima (i.e., local minima 3 and 4), and $l^{\rightarrow Hazardous}(s_k)$ represents the shortest path length from state $s_k$ to hazardous minima (i.e., local minima 1, 2 and 5). Considering that some states cannot reach either hazardous or normal minima, we set the corresponding value of $l^{\rightarrow Normal}(s_k)$ or $l^{\rightarrow Hazardous}(s_k)$ to 100, which is significantly greater than all shortest path lengths towards a certain type of minima (Supplementary Fig. 5). According to this definition, a high *R*-value indicates short path-length towards hazardous minima, signifying great risk.

## Acknowledgments

This work was supported by the National Natural Science Foundation of China (Grants 72225012, 72288101, and 71822101), the Fundamental Research Funds for the Central Universities.

## Data Availability

The data that support the findings of this study are available on request from corresponding authors.

## Code Availability

The code is available upon request directly from corresponding authors.

## Additional information

More details and figures are presented in Supplementary information.

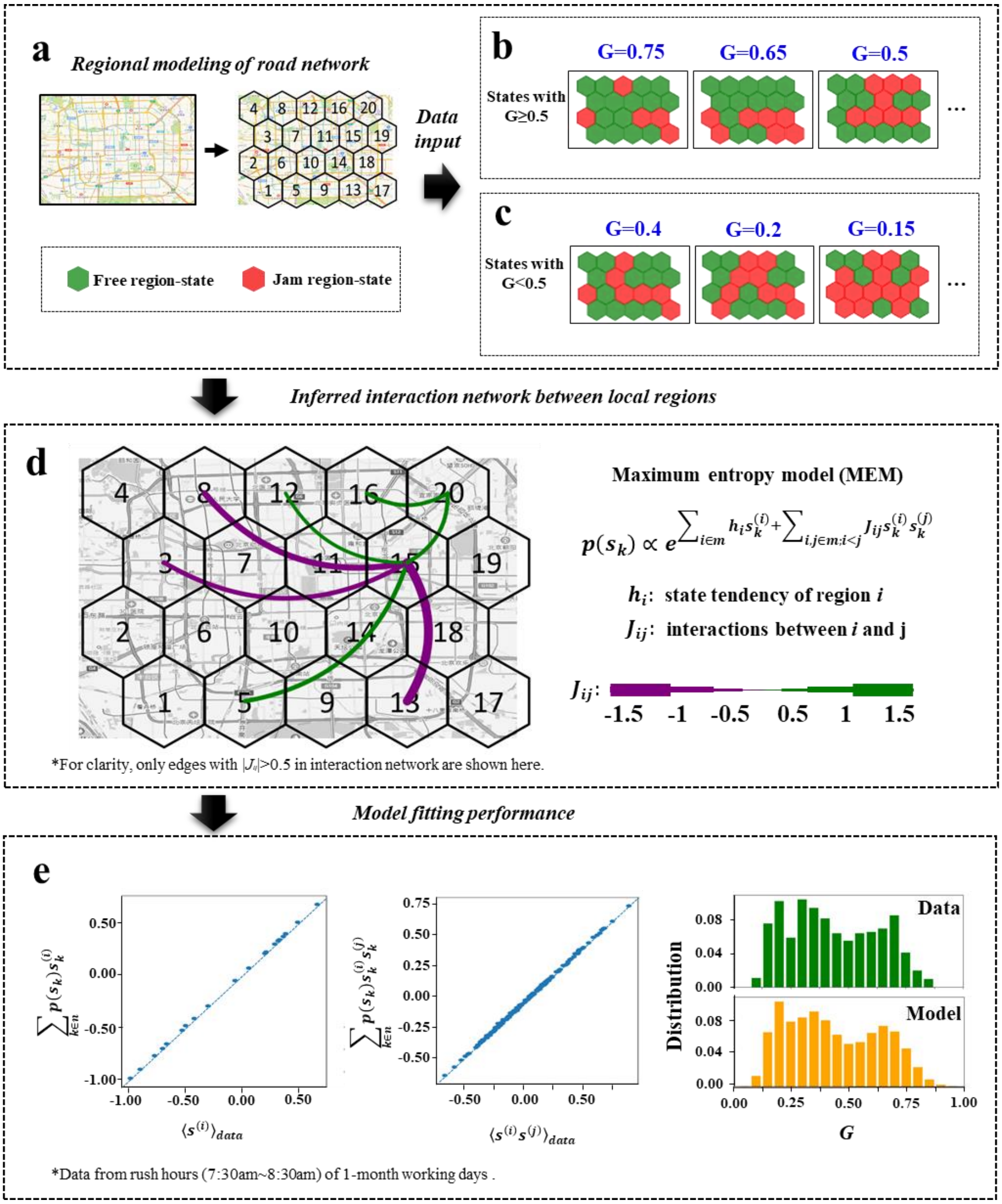

**Fig1. Pairwise maximum entropy model for urban traffic. (a) Regional modeling of Beijing road network (within the 4th Ring Road)**. The entire road network is divided into 20 regular hexagonal local regions of equal size. Local region state can be either free (i.e., green) or jam (i.e., red) based on the corresponding local road network (see Methods). Traffic functional performance of network states ($G$) is defined as the ratio of largest functional cluster[33] formed by connected free regions to the total number of regions. **(b) Typical network states with G ≥ 0.5 observed in real data**. These states represent good traffic operational connectivity. **(c) Typical network states with G < 0.5 observed in real data**. These states represent poor traffic operational connectivity. **(d) Inferred interaction network between local regions based on pairwise maximum entropy model**. Pairwise maximum entropy model parameters (i.e., $h_i$ and $J_{ij}$) are estimated from real data. **(e) Model goodness**. The fitting performance of the model to rush hour data (7:30am~8:30am) is good (see main text for details).

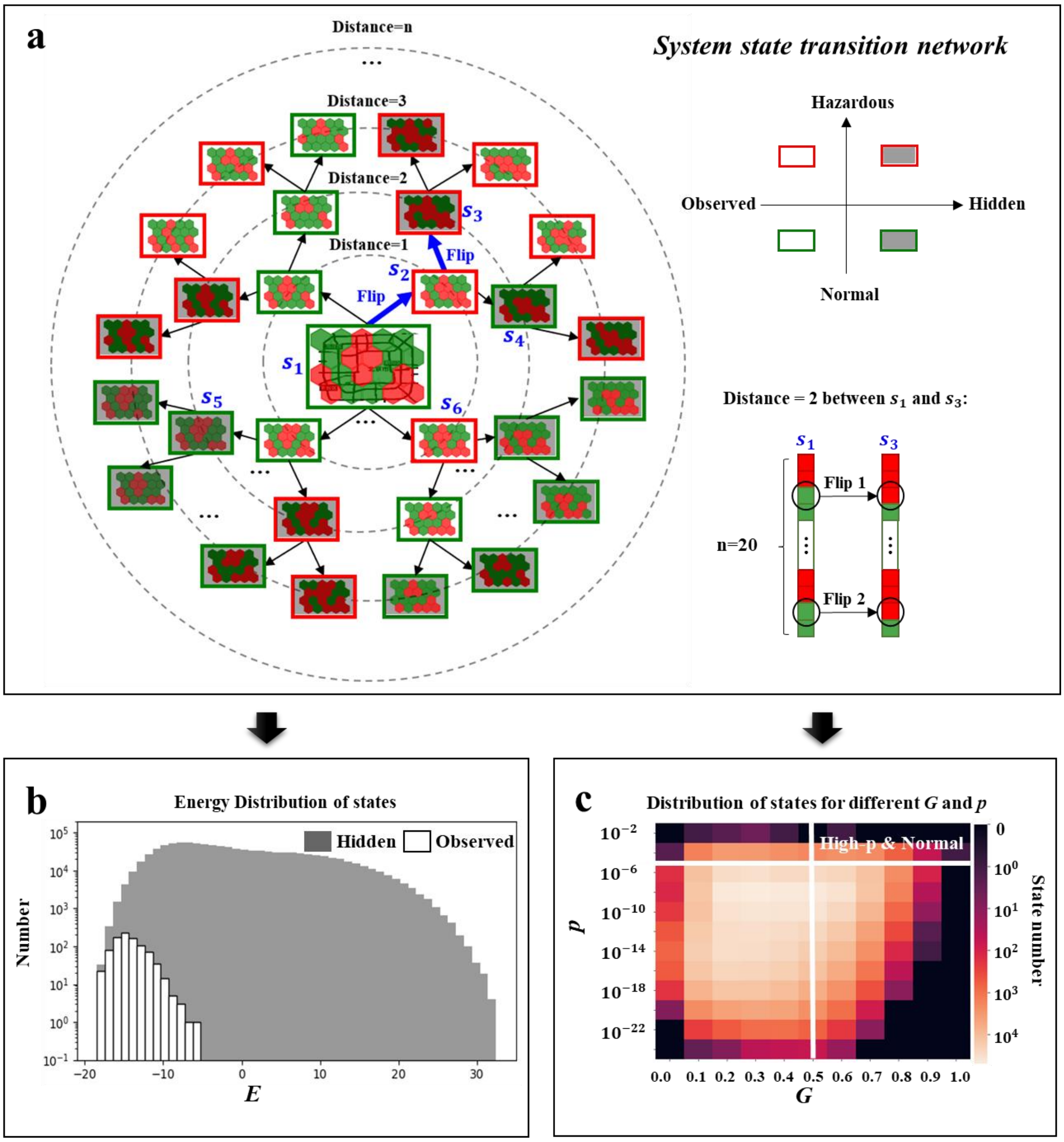

**Fig2. State space analysis based on pairwise maximum entropy model. (a) System state transition network**. Nodes in this network represent possible states of entire road network, categorized into 4 classes: hidden hazardous states (e.g., $s_3$), hidden normal states (e.g., $s_4$), observed hazardous states (e.g., $s_2$) and observed normal states (e.g., $s_1$). We define distance between two states as number of region-state changes required to transition between them, e.g., distance between $s_3$ and $s_1$ equals 2. **(b) Energy distribution of states.** Energy values of all network states are distributed over a wide range between -20 and 30. While a few observed states concentrate around -15, there are still a fraction of hidden states with low energy that have never been observed in real data. **(c) Distribution of states for different *G* and *p*.** High-p states are defined as states with $E < -13.2365$ (i.e., $p > 1\times10^{-5}$), distributed over different values of G. Hidden high-risky states are confined within these high-p and normal states (i.e., $G \geq 0.5$).

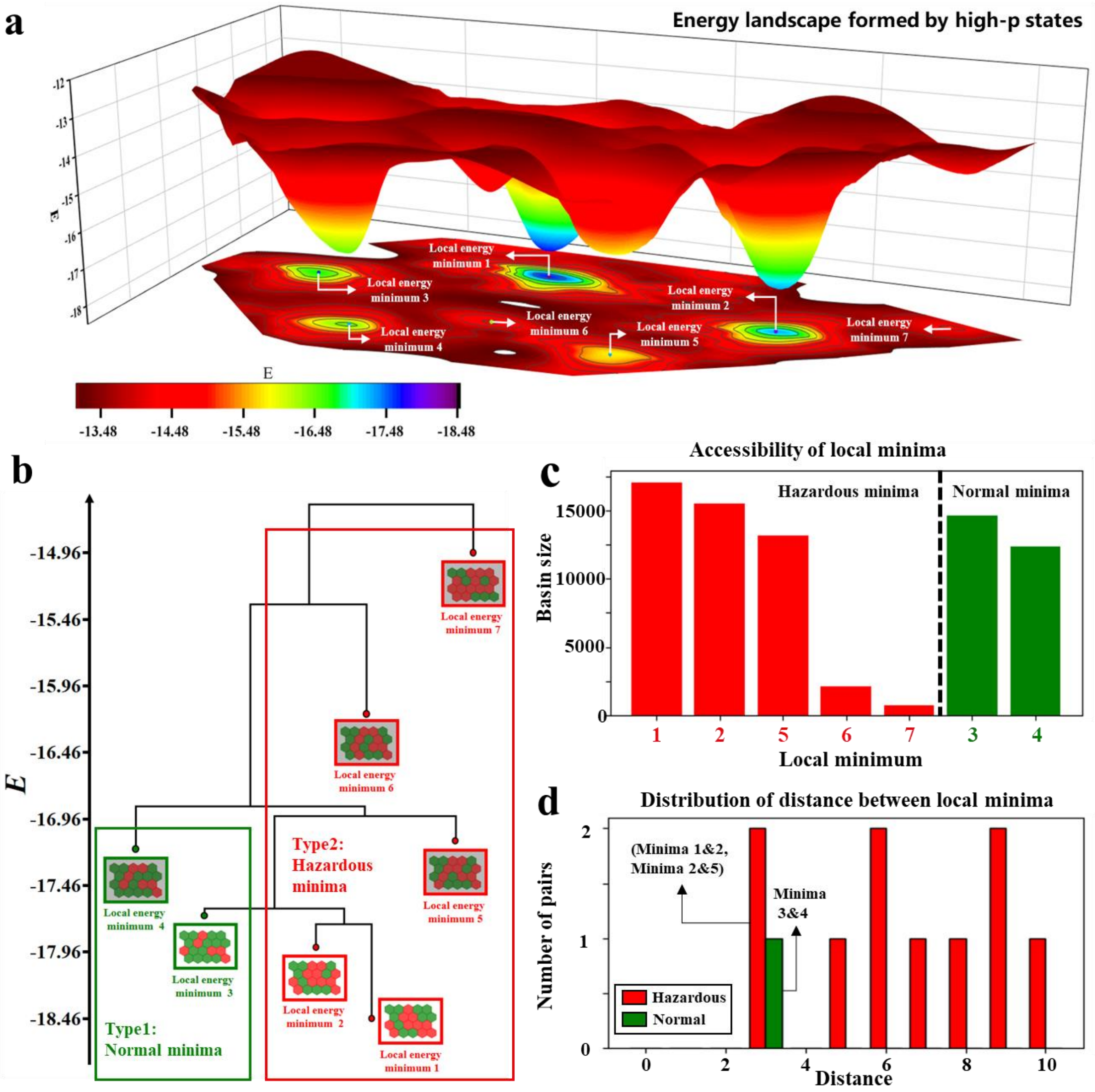

**Fig3. System energy landscape. (a) Energy landscape formed by high-p states.** To emphasize local energy minima, visualization of energy landscape simplifies dynamical processes of states (see Methods) and relative position inside each minimum basin is only for illustration. **(b) Disconnectivity graph of energy landscape.** It illustrates energy levels of local energy minima and energy barriers between them. There are seven energy minima with energy values ranging between -14 and -19, and two types of local minima, i.e., normal ($G \geq 0.5$) and hazardous ($G < 0.5$). **(c) Accessibility of local energy minima**. Compared to normal minima, basin size of hazardous minima is relatively larger. **(d) Distribution of distance between local minima.** Compared to distances between other hazardous minima, the distances between minima 1 and 2, as well as between minima 2 and 5, are shorter.

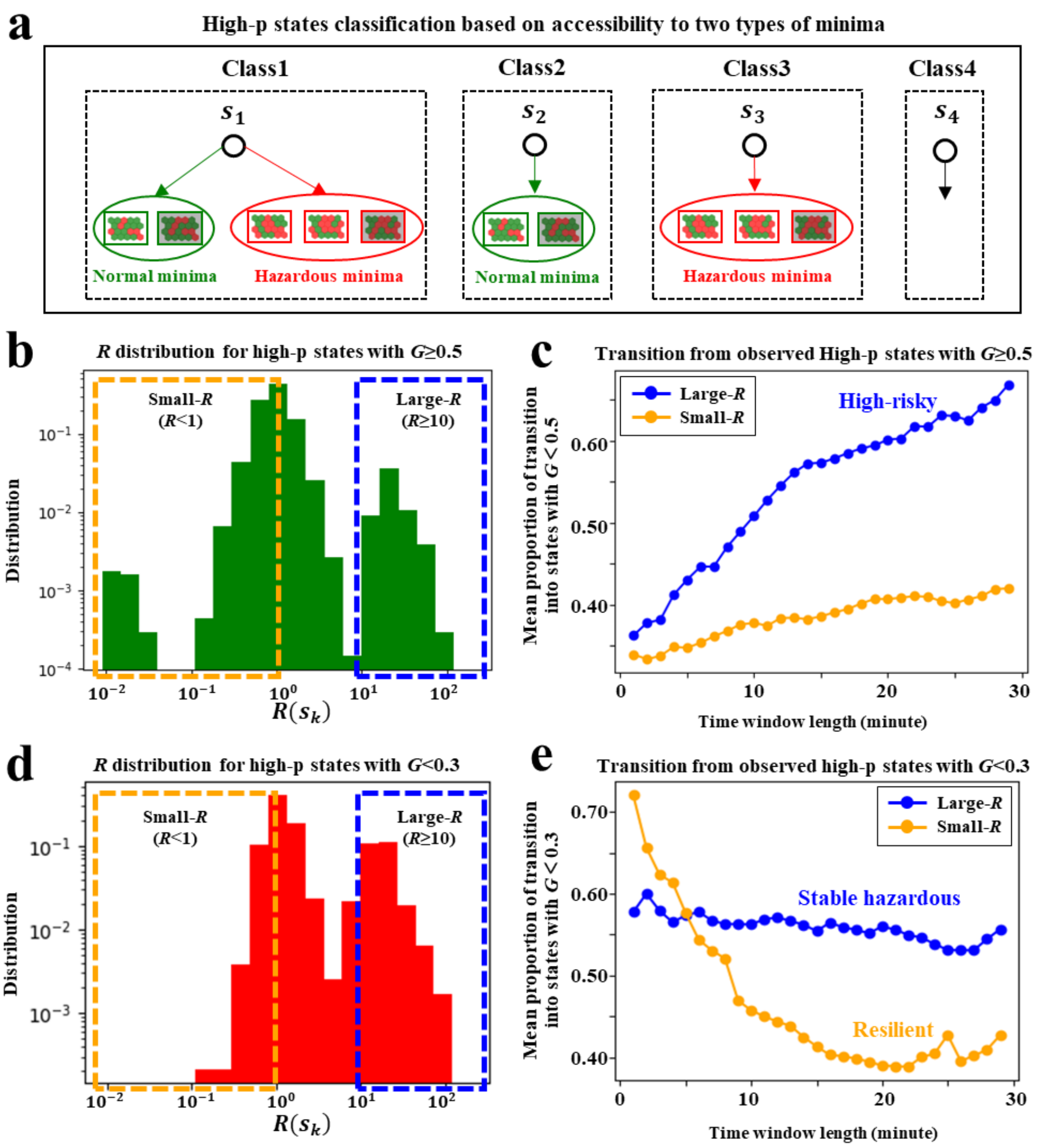

**Fig4. Hidden high-risky states identification. (a) High-p states classification based on accessibility to two types of minima.** There are four classes of high-p states based on their accessibility to normal minima and hazardous minima: states that can reach both two types of minima (class 1), states that can only reach normal minima (class 2), states that can only reach hazardous minima (class 3), and states that are inaccessible to both two types of minima (class 4). **(b)** ***R*** **distribution for high-p states with G≥0.5.** There is a broad distribution of *R*-value ranging from 0.01 to 100 for high-p states with G≥0.5. **(c) Actual dynamical processes transitioning from observed high-p states with G≥0.5.** Compared to observed normal (i.e., G≥0.5) high-p states with small-*R*, observed normal high-p states with large-*R* are more likely to transition into hazardous states (i.e., G<0.5). **(d)** ***R*** **distribution for high-p states with** ***G*****<0.3.** There is a broad distribution of *R*-value ranging from 0.1 to 100 for high-p states with G<0.3. **(e) Actual dynamical processes transitioning from observed high-p states with** ***G*****<0.3.** High-p states with *G*<0.3 and small-*R* possess an inherent ability to spontaneously escape such states, highlighting system resilience. Conversely, high-p states with *G*<0.3 and large-*R* tend to lead the system into long-time poor traffic operational connectivity, which can be regarded as stable hazardous states.